\documentclass{appolb}
\usepackage{graphicx}
\usepackage{amssymb}
\usepackage{amsmath}
\usepackage[table]{xcolor}
\usepackage{float}
\usepackage{soul}
\usepackage{color}

\begin{document}
\title{A RANKING METHOD FOR SELECTION OF $\eta$ MESONS IN HIGH MULTIPLICITY EVENTS}

\author{A.~Bing\"ul, U.~\c Sa\c smaz
	\address{Gaziantep University, Dep. of Engineering Physics, Gaziantep 27310, Turkey}
	\\ 
	{~}
	\\
	A.~J.~Beddall
	\address{Bah\c{c}e\c{s}ehir University, Faculty of Engineering and Natural Sciences, Istanbul, Turkey}
}

\maketitle
\begin{abstract}
The selection of $\eta$ mesons with a high efficiency and a high purity
can be important in the formation of statistically 
significant invariant mass spectra in the reconstruction of short-lived particles 
such as $\eta' \rightarrow \pi^{+} \pi^{-} \eta$. 
In this study, a cut-based standard method and a Ranking method 
to reduce combinatorial background in the reconstruction of 
$\eta \rightarrow \gamma \gamma$ decays in high multiplicity hadronic 
events are presented. By using recorded ALEPH data and fully simulated events, 
the performances of the methods are compared. 
Results show that the Ranking method yields significant improvements in the purity 
of the selected $\eta$ meson relative to the standard method.
\end{abstract}
  
%%%%%%%%%%%%%%%%%%%%%%%%%%%%%%%%%%%%%%%%%%%%%%%%%%%%%%%%%%%%%%%%%%%%%%%%%%%%%%%%%%%%%%%%%%%%%%%%%%%%%
\section{Introduction}
\label{SECTION_INTRO}

In particle collisions at high energies, one can have  
events containing a high multiplicity of hadronic particles.
An example is the production of $Z$ Bosons 
from the decay channel $Z \rightarrow q \bar{q}$ at LEP,
the $e^- e^+$ collider at $\sqrt{s} = 91.2$ GeV.
Two or more jets of hadrons and other particles,
on average 21 charged and 21 neutral particles per event,
are produced by the hadronization of quarks and
gluons in such decays~\cite{CITE_PDG}.

Invariant mass spectra, built from selected
final state particles, can reveal the presence of short-lived
`mother' particles which present themselves as
peaks on a combinatorial background.
An example decay is 
$\eta'(958) \rightarrow \pi^{+} \pi^{-} \eta$.
Here, the $\eta'$ is reconstructed from measurements of the
momenta of charged pions with good momentum resolution in the 
tracking chambers, and the $\eta$ meson can be reconstructed from  
the two-photon channel $\eta(548) \rightarrow \gamma \gamma$
where the daughter photons are measured in the electromagnetic calorimeter 
typically with relatively poor energy resolution.

The selection of $\pi^0$ and $\eta$ mesons in the two photon decay 
mode can be a relatively simple task where 
candidates are selected from a mass window around the signal peaks. 
However, in environments where  particle multiplicities are high,
such as at LEP and LHC, further analysis and optimization can result 
in a higher selection purity and efficiency.

In this study, two methods for improving selection purity of $\eta$ mesons
are investigated in detail; a standard method and a Ranking method.
First, the standard method which is used to select signal candidates from a mass window
around the $\eta$ peak and to reject photons from neutral pion decays is described.
Then, a probability density estimator, 
based on reconstructed kinematic parameters of the $\eta$ mesons, 
for distinguishing backgrounds and signals, and
the Ranking method used for further improvement in the 
purity of reconstructed $\eta$ mesons are 
presented respectively.
Finally, using ALEPH Archived Data and Simulation~\cite{CITE_ALEPH_RULE},
example applications of these two $\eta$ selection methods for improving signal 
significance of the decay channel $\eta' \rightarrow \pi^{+} \pi^{-} \eta$ 
are demonstrated at the end.

%%%%%%%%%%%%%%%%%%%%%%%%%%%%%%%%%%%%%%%%%%%%%%%%%%%%%%%%%%%%%%%%%%%%%%%%%%%%%%%%%%%%%%%
\section{Event and track selection}
\label{SECTION_EVENT}

In the event simulation, the decays 
$\pi^0 \rightarrow \gamma \gamma$,
$\eta \rightarrow \gamma \gamma$,
and
$\eta' \rightarrow \pi^+ \pi^- \eta$
are selected from $e^- e^+$ collision events representing 
simulations of hadronic $Z$ decays at the LEP collider. 
The selected events are passed through the full simulation
and the reconstruction program for the ALEPH
detector~\cite{CITE_ALEPH} and so provide a realistic
simulation of daughter momentum resolutions.
Totally, 4,923,816 reconstructed events are used in the simulation
studies.

Using the hadronic event selection criteria described in~\cite{CITE_ALEPH_EVENT},
a total of 3,239,746 hadronic $Z$ decays around
$\sqrt{s} = 91.2$ GeV recorded by ALEPH at LEP in the period between 
1991 and 1995 are selected for the real data analysis.

In the physics analysis, unconverted photons with an energy greater 
than $0.8$ GeV are selected.
The reconstructed charged particles are required to have a 
polar angle in the range $20^\circ < \theta <160^\circ$ and a
transverse momentum of at least $0.5$ GeV.

%%%%%%%%%%%%%%%%%%%%%%%%%%%%%%%%%%%%%%%%%%%%%%%%%%%%%%%%%%%%%%%%%%%%%%%%%%%%%%%%%%%%%%%
\section{Standard method}
\label{SECTION_STANDARD}

\begin{figure*}[!htbp]
\centering
\begin{center}
	\resizebox{0.8\textwidth}{!}{\includegraphics{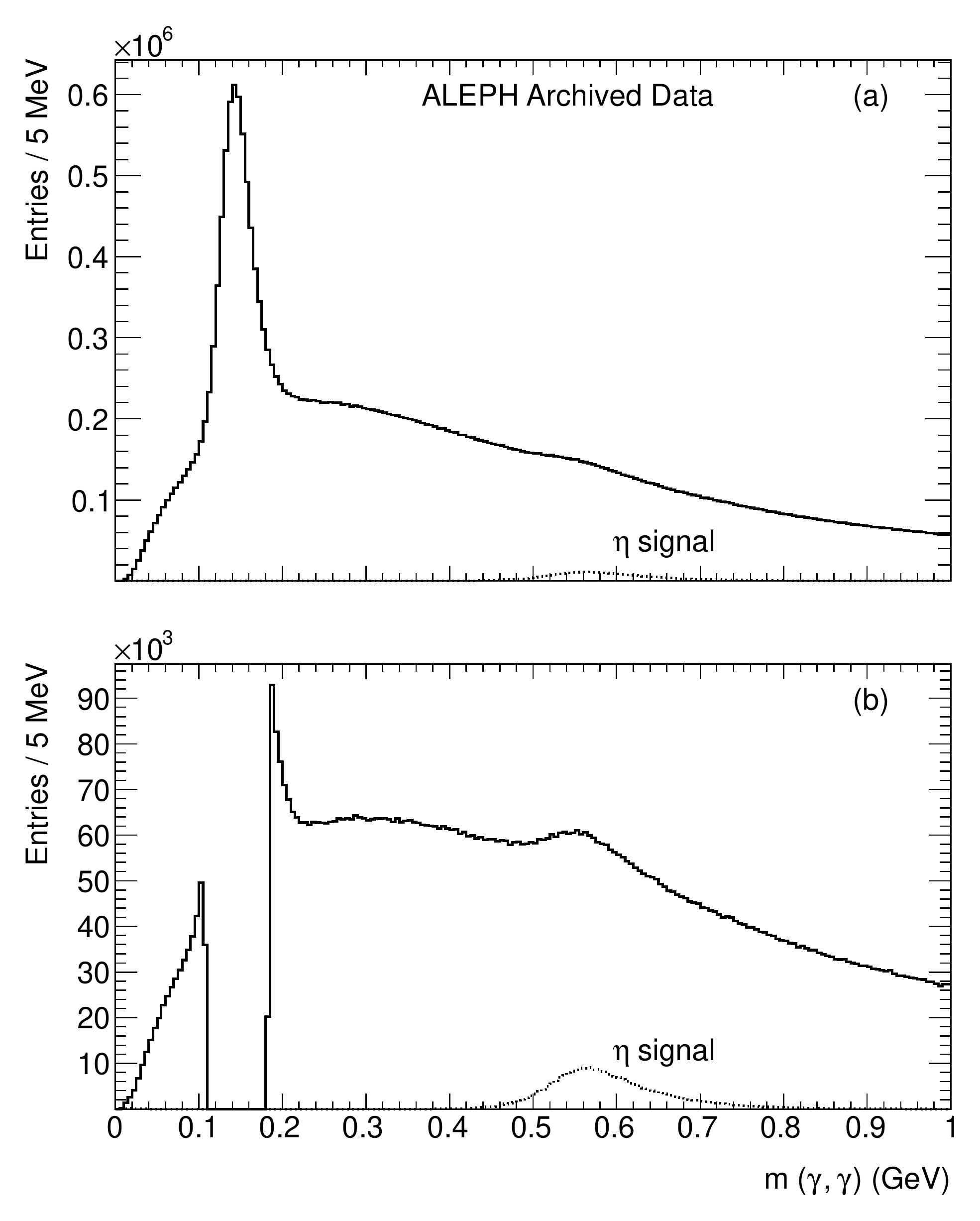}}
	\vspace{-0mm}
	\caption{\label{FIGURE_MGG}
		Example of a two-photon invariant mass distribution (a) before and (b)
		after $\pm 2 \sigma_r$ rejection around the $\pi^0$ signal in simulation.
		The $\eta$ signal is also shown by a dotted line under the full mass spectra.
		It is clear that the neutral pion rejection improves 
		$\eta$ signal significance by reducing combinatorial background 
		under the $\eta$ signal in case (a).
	}
\end{center}
\end{figure*}

$\pi^0$ and $\eta$ candidates are reconstructed by combining pairs of photons.
The branching ratios of the decays $\pi^0  \rightarrow \gamma \gamma$ and 
$\eta \rightarrow \gamma \gamma$ are about 98.8\% and 39.4\%, respectively~\cite{CITE_PDG}.
The $\pi^0$ signal around 0.135 GeV and $\eta$ signal around 0.548 GeV 
can be seen in two-photon invariant mass spectra, as shown in Figure~\ref{FIGURE_MGG}a. 
Due to its smaller mass and higher branching ratio, 
the $\pi^0$ multiplicity is much greater than that of the $\eta$; 
the $\pi^0$ and its combinatorial background therefore dominate
the two-photon mass spectra making the $\eta$ signal difficult
to distinguish from the large combinatorial background.

$\pi^0$ and $\eta$ candidates are directly selected from a mass
window around the signal peaks. The selection of these particles with a 
high purity and high efficiency is important. 
In this study, we define the purity ($P$) and efficiency ($E$) as follows:

\begin{equation}
P = S / (S+B)
\end{equation}

\begin{equation}
E = S / S_0
\end{equation}
where $S$ is the number of signal entries, $B$ is the number of background entries 
within the given mass window and $S_0$ is the total number of 
signal entries within $\pm 6 \sigma$ mass window.
Here, $\sigma$ is the mass resolution which, in this study, is defined as one 
standard deviation of the signal mass distribution in the simulation.

Note that, in the photon matching, a search for the best match between 
the generated and reconstructed photon candidates is performed
based on the angular distance between reconstructed and truth photons.
%\footnote{Here, 
%	$\Delta \eta_r$ and $\Delta \phi$ are the differences of pseudo-rapidities and 
%	azimuthal angles respectively between reconstructed and truth photons.}
%defined by 
%$\Delta R = \sqrt{\Delta \eta_r^2 + \Delta \phi^2}$.
A pair with the smallest angular distance is considered to be the best match.
To define the signal, $S$, two matched photons are combined if they originate 
from the same parent ($\eta$ or $\pi^0$).

The poor purity of the $\eta$ signal can be improved by rejecting photons 
that appear to originate from a $\pi^0$ decay ($\pi^0$ rejection).
This is achieved by eliminating the photon pairs with an invariant mass within
$\pm p \sigma_r$ rejection mass window around the $\pi^0$ peak where $\sigma_r$ is the $\pi^0$
mass resolution and $p$ is a real number.
Figure~\ref{FIGURE_MGG}b shows the invariant mass spectra after $\pm 2\sigma_r$ rejection
around $\pi^0$ signal. The $\eta$ peak, though significantly diminished, is much clearer 
due to the greatly reduced background after $\pi^0$ rejection.

After neutral pion rejection, $\eta$ candidates can be selected from a mass window of 
size $\pm q \sigma_s$ around the $\eta$ peak. Here, $\sigma_s$ is the $\eta$ signal 
mass resolution, and $q$ is a scale that directly effects the selection purity and 
efficiency. A narrow window (small $q$) will increase purity by
selecting less background but reduce efficiency as it selects less signal.
A narrow selection window will also tend to increase systematic errors
when the mass spectra of data and simulation are not in good agreement.
The optimal value of $q$ is therefore a balance between purity and efficiency.
We define the optimization condition such that the product
$E \times P$ is maximum\footnote{The product $E \times P$ is closely related to signal to
	noise ratio which is defined as $S/N = S/\sqrt{S+B}$. Thus $E \times P = (S/N)^2/S_0$.}. 

This procedure was applied in Ref~\cite{CITE_ALEPH_ETA} in order to
extract the production rates of $\eta$ and $\eta^\prime$ in hadronic Z decays
by the ALEPH Collaboration using data collected during the 1990 and 1995 running period of LEP.
They preferred the mass window scales of about $p=1.5$ and $q=1.5$
resulting in $P = 9.6\%$, $E = 71.0\%$ and $E\times P = 6.8\%$.
However, in this study, the best scales are
found to be $p = 2.0$ and $q = 1.5$, and the corresponding values for purity and 
efficiency are $P = 10.2\%$ and $E = 67.4\%$ respectively giving the optimal product 
$E\times P = 6.9\%$. Although the statistical performance is very similar in the two
cases, the wider selection window used in this study
would favour lower systematic errors.
The effect of varying $p$ and $q$ values will be discussed in Section~\ref{SECTION_RANKING}.

In this study, the $\pi^0$ rejection and $\eta$ selection procedure 
defined above is called the `standard method' which forms the starting
point to study another method for improving the purity of $\eta$ mesons. 

%%%%%%%%%%%%%%%%%%%%%%%%%%%%%%%%%%%%%%%%%%%%%%%%%%%%%%%%%%%%%%%%%%%%%%%%%%%%%%%%%%%%%%%
\section{$\eta$ estimator}
\label{SECTION_ESTIMATOR}
An estimator obtained from kinematic properties of the $\eta$ signal can be defined to 
discriminate between the $\eta$ signal and background. In this study, three discriminating variables 
are used as the $\eta$ estimator; the opening angle between photon pairs ($\theta_{12}$),
invariant mass of the pairs ($M_{12}$) and total energy of the pairs ($E_{12}$).

\begin{figure*}[!htbp]
\centering
\begin{center}
	\resizebox{1.0\textwidth}{!}{\includegraphics{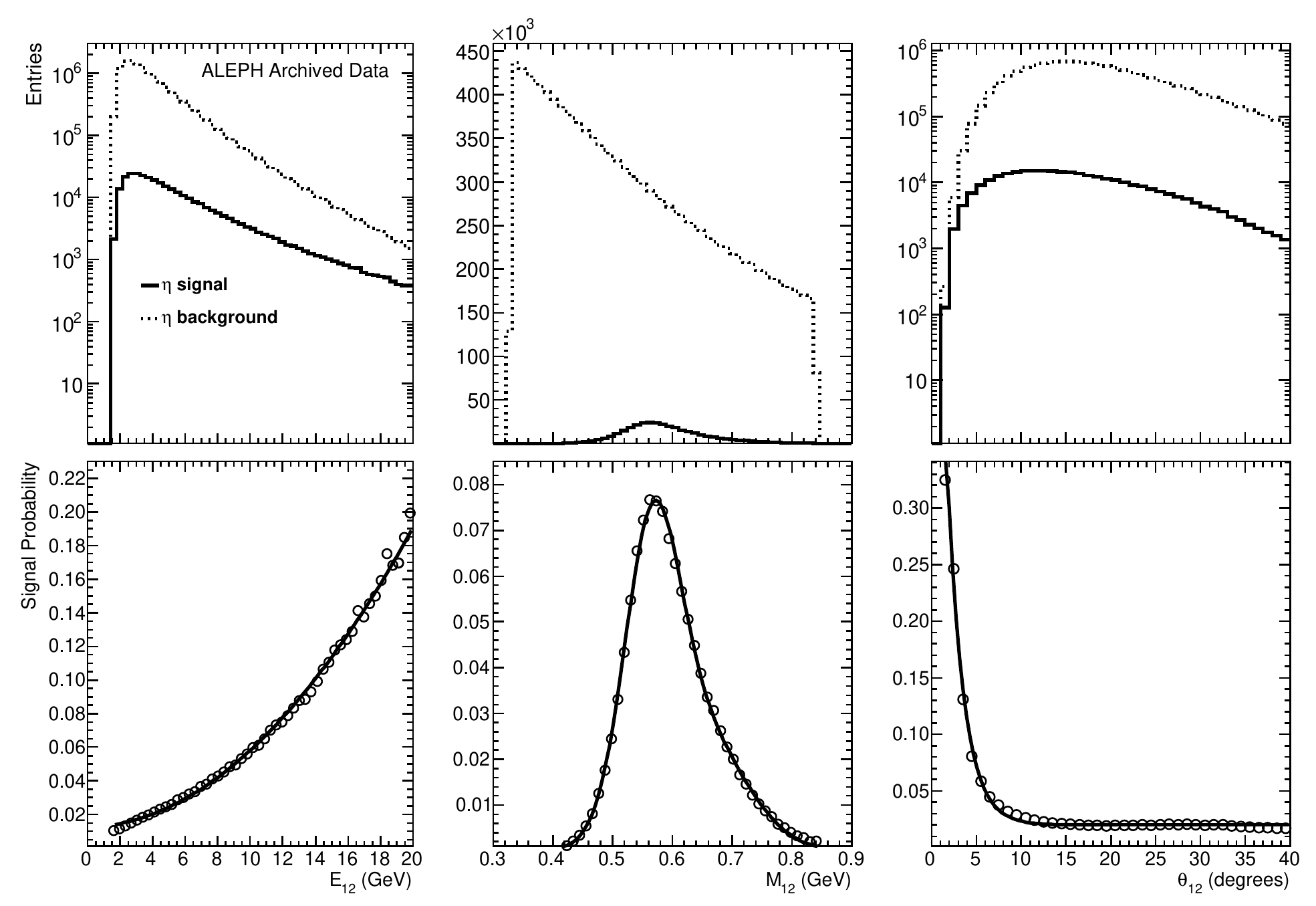}}
	\vspace{-0mm}
	\caption{\label{FIGURE_VARIABLES}
		Distributions of discriminating variables used in the study 
		(top row) and their corresponding purity distributions (bottom row)
		indicated by open circles in simulation. For each variable, 
		the purity histogram is obtained from the ratio of the signal 
		(solid line) to the sum of signal and background (dotted line) 
		histograms. Each purity histogram is fitted with a suitable 
		function such that $\chi^2 / \text{ndf}<1$.
	}
\end{center}
\end{figure*}

The top row of Figure~\ref{FIGURE_VARIABLES} shows the distribution of these variables in
simulation for $p=0$ and $q=4$. 
The signal (solid line) and background (dotted line)
components are shown separately. In the bottom row, for each variable, a purity histogram is 
obtained from the ratio of the signal to the sum of signal and background histograms. 
Each purity histogram is then fitted with a suitable function.
It is clear that the signal tends to have smaller values of opening angles,
closer mass values to the nominal mass (0.548 GeV) and larger energy values
than the background.

%To represent the statistical relationship among these discriminating variables, 
%the Pearson product-moment correlation coefficients~\cite{CITE_CORRELATION},
%$\rho$, are calculated.
%The values in MC study are obtained as $\rho(\theta_{12}, M_{12})=0.55$, 
%$\rho(\theta_{12}, E_{12})=-0.63$ and $\rho(E_{12}, M_{12})=0.04$. 
%Thus, the discriminating variables are not strongly correlated since
%the correlation coefficients are not close to $\pm 1$. As a result, 
A probability density estimator (PDE) for $\eta$ candidates can be built from the product 
of these three purity (or probability) values evaluated from the best fit functions.
Hence the PDE function is given by:

\begin{equation}
\label{EQN_PDE}
e_{PDE}(\theta_{12}, M_{12}, E_{12}) = f_1(\theta_{12}) f_2(M_{12}) f_3(E_{12})
\end{equation}
where the explicit forms of the fit functions are respectively as follows:

\begin{align}
f_1(\theta_{12}) &= a_0+a_1 e^{a_2 \theta_{12}} \\
f_2(M_{12})      &= a_3 e^{-((M_{12}-a_4)/a_5)^2 /2} + a_6 e^{-((M_{12}-a_7)/a_8)^2/2} \\
f_3(E_{12})      &= a_9 + a_{10} E_{12} + a_{11} E_{12}^2
\end{align}
Here, $a_i$ are free parameters obtained from the fitting procedure.

\begin{figure*}[!htbp]
\centering
\begin{center}
	\resizebox{0.8\textwidth}{!}{\includegraphics{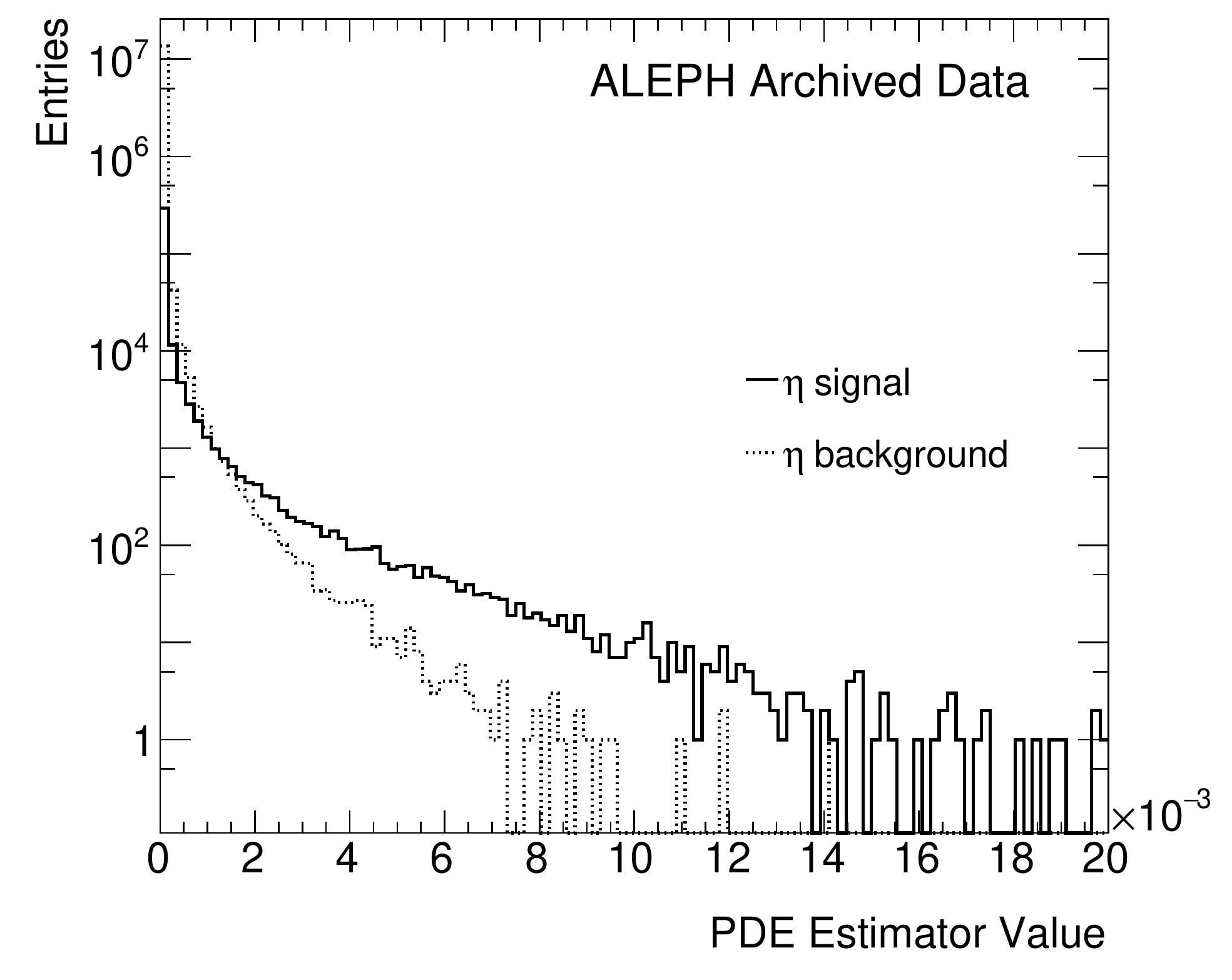}}
	\vspace{-0mm}
	\caption{\label{FIGURE_ESTIMATOR}			
		Distributions of estimator values obtained from 
		the PDE method for discriminating variables, 
		(Eqn~\ref{EQN_PDE}), for the $\eta$ signal (solid line) and 
		the background (dotted line) components.
		Candidates are selected from a large window ($q=4$) 
		with no pion rejection ($p=0$).
	}
\end{center}
\end{figure*}

Figure~\ref{FIGURE_ESTIMATOR} shows the distribution of the PDE values
for the signal (solid line) and the background (dotted line). 
As expected, $\eta$ candidates that are formed from $\eta$ decays (the signal)
tend to have larger PDE values than $\eta$ candidates formed from the combinatorial background.
Equation~\ref{EQN_PDE} can therefore be used to attempt to distinguish 
between correct and wrong combinations of photon pairs.

\begin{figure*}[!htbp]
\centering
\begin{center}
	\resizebox{0.8\textwidth}{!}{\includegraphics{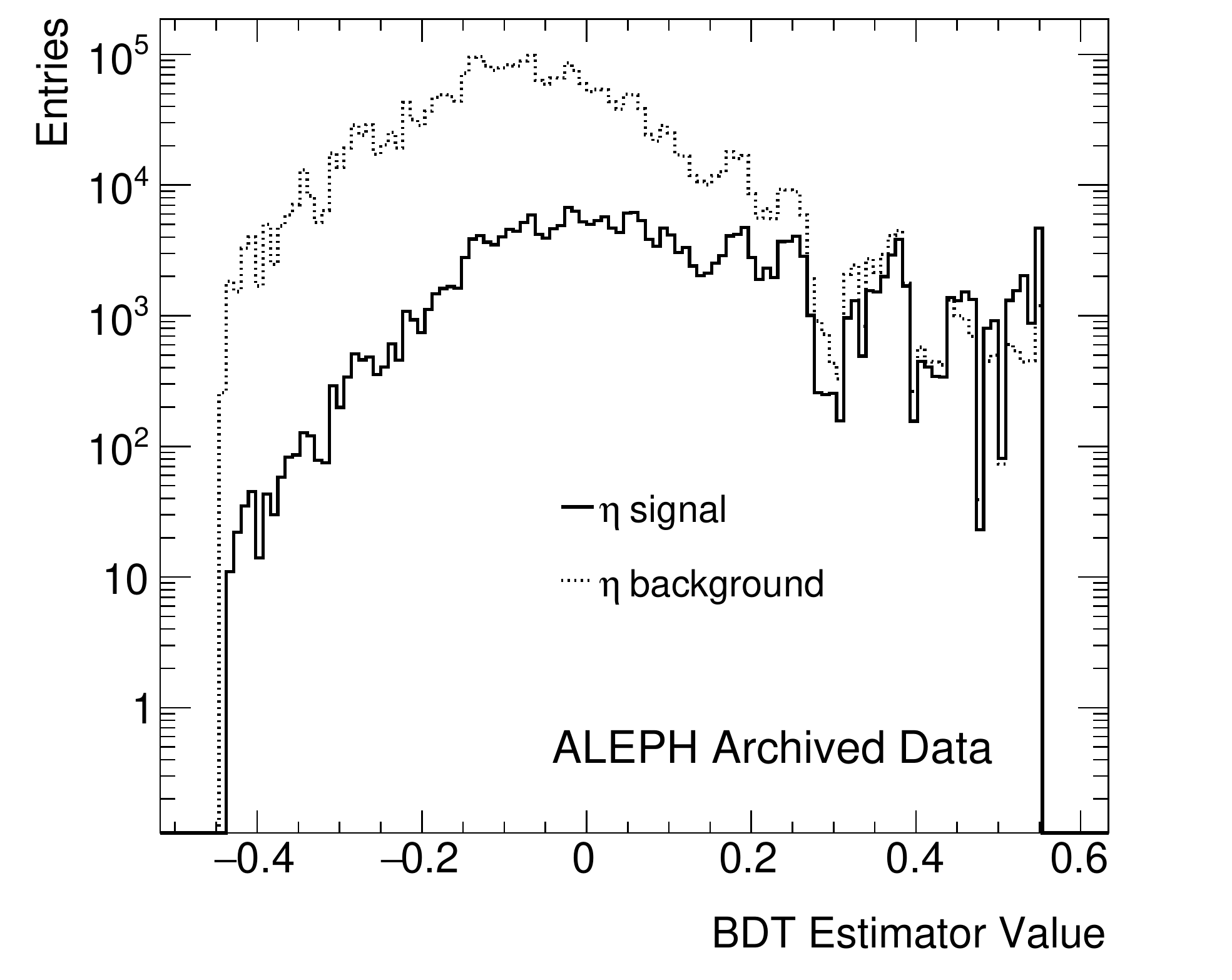}}
	\vspace{-0mm}
	\caption{\label{FIGURE_ESTIMATOR_BDT}			
		Distributions of estimator values obtained from 
		the BDT method for discriminating variables. 
		Candidates are selected from a large window ($q=4$) 
		with no pion rejection ($p=0$).
	}
\end{center}
\end{figure*}

%Demonstrations of signal and background discrimination often include a comparison
%with results from boosted decision trees or neural networks since they generally 
%give better performance. 
Advanced multivariate classifiers such as boosted decision trees and neural networks
have become increasingly popular in particle physics
due to their superior discrimination capabilities and growth
of available computing power in recent years.
In order to compare with the PDE, we use the boosted 
decision trees (BDT) method with 600 
trees\footnote{600 trees are found to be optimum to reduce background candidates maximally
	while saving more signal.} 
in the Toolkit for MultiVariate 
data Analysis with ROOT (TMVA)~\cite{CITE_ROOT}.
The distribution of estimator values obtained from the BDT method 
is shown in Figure~\ref{FIGURE_ESTIMATOR_BDT}. $\eta$ signals, on average, 
have larger estimator values than background as in the PDE method. 

Note that, the compact analytical form of the estimator function for this case, 
$e_{BDT}(\theta_{12}, M_{12}, E_{12})$, cannot be written
explicitly as in Equation~\ref{EQN_PDE}. 
Instead it is evaluated using the parameterized output file obtained from BDT training.
It is also clear that BDT (and neural networks) can capture dependencies 
between the variables while the PDE method implicitly assumes
that the variables are independent.

%%%%%%%%%%%%%%%%%%%%%%%%%%%%%%%%%%%%%%%%%%%%%%%%%%%%%%%%%%%%%%%%%%%%%%%%%%%%%%%%%%%%%%%
\section{Ranking method}
\label{SECTION_RANKING}
Further improvement in $\eta$ purity is achieved by 
applying a Ranking method to the remaining candidates after 
the initial mass window cuts. 
The Ranking method with a relatively crude estimator 
and its performance for 
the neutral pion selection are described in detail 
elsewhere~\cite{CITE_RANKING}\footnote{In Ref~\cite{CITE_RANKING},
	only simulated events were used to generate hadronic decays of
	the $Z$ boson at various center of mass energies.}.
In this paper, the method is improved by using more sophisticated 
estimators and applied to the $\eta \rightarrow \gamma \gamma$ channel for the first time. 
A summary of the method is as follows:

\begin{enumerate}
	\item An estimator value is assigned for each
	$\eta$ candidate in an event according to the 
	values from the estimator function
	defined in Section~\ref{SECTION_ESTIMATOR}.
	\item Candidates are then ranked in decreasing order of estimator 
	(true $\eta$s are most likely to be nearer the top of the list).
	\item Candidates are removed from the list if their 
	estimator values are less than a predefined value of an estimator cut, $e_{cut}$
	(for this study, the optimal values corresponding to maximum $E \times P$ 
	are found to be $e_{cut} = 2.5 \times 10^{-5}$ for the PDE method and $e_{cut} = 0.0$
	for the BDT method).
	\item A scan is then made through the list for pairs of 
	$\eta$ candidates which share photons. If there exists such a pair, 
	the candidate with the smallest estimator value is removed from the list.
\end{enumerate}

An example application of the Ranking method in a simulated event containing 
ten $\eta$ candidates selected for $p = 2$ and $q = 4$
is shown below. $\eta$ candidates are labeled with integers from 1 to 10
and photons are represented by integers from 21 to 29.
\\ \\
\noindent Candidates before Ranking:

\begin{center}
\begin{tabular}{l l l l l}
	\hline
	$\eta$ signal    & Estimator & Photon   &   &  Truth   \\
	Candidate & Value     & \#1      & \#2 &  Info.   \\
	\hline
	1        &  1.313e-03 & 22 & 24 & True  \\
	2        &  4.148e-04 & 26 & 28 & False \\
	3        &  3.381e-04 & 22 & 25 & False \\
	4        &  1.276e-04 & 21 & 28 & False \\
	5        &  1.252e-04 & 23 & 25 & False \\
	6        &  3.726e-05 & 23 & 26 & True  \\
	7        &  2.535e-05 & 23 & 29 & False \\
	8        &  1.831e-05 & 21 & 23 & False \\
	9        &  8.132e-06 & 24 & 25 & False \\
	10        &  5.654e-06 & 22 & 29 & False \\
	\hline 
\end{tabular}

\end{center}

According to the Ranking algorithm above, the last three candidates in the
list must be removed since their estimator values
are less than $e_{cut} = 2.5 \times 10^{-5}$. $\eta$ candidate 1
removes candidate 3 from the list as they both share
photon 22. Similarly, candidate 2 removes candidate 4,
and candidate 5 removes candidate 6 and 7. Although the 6th candidate 
is a {\it signal}, it is removed from the list since it shares the photon 23
with a background (candidate 5) which has, by chance, a larger estimator value.
As a result, the list of selected candidates after Ranking is as follows: \\

\begin{center}
\begin{tabular}{l l l l l}
	\hline
	$\eta$ signal    & Estimator & Photon   &   &  Truth   \\
	Candidate & Value     & \#1      & \#2 &  Info.   \\
	\hline
	1        &  1.313e-03 & 22 & 24 & True  \\
	2        &  4.148e-04 & 26 & 28 & False \\
	5        &  1.252e-04 & 23 & 25 & False \\
	\hline
\end{tabular}
\end{center}

After Ranking, one out of two true candidates are selected,
whereas six out of eight background candidates are rejected.
On average, Ranking improves the selection purity of $\eta$ mesons
with a reduction in selection efficiency.

%%%%%%%%%%%%%%%%%%%%%%%%%%%%%%%%%%%%%%%%%%%%%%%%%%%%%%%%%%%%%%%%%%%%%%%%%%%%%%%%%%%%%%%
\section{Performance}
\label{SECTION_PERFORMANCE}
To investigate the performance of the Ranking
method and the standard method, the width of $\pi^0$ rejection
window (controlled by the scale $p$) and the width of $\eta$ selection mass
window (controlled by the scale $q$) are varied.
The selection efficiency and purity, and their product, are then used
to compare the performance of the two methods.
Note that, the selection efficiency of $\eta$ signal is calculated 
for $p=0$ and $q=6$.

Detailed results of the study is shown in Table~\ref{TABLE_EFFPUR}.
For both methods, improvements in purity and the product of purity and efficiency 
are evident with respect to $p=0$.
However, the Ranking method using either PDE or BDT significantly improves the 
selection purity of $\eta$ signals compared to the standard method for all values of $p$ and $q$.
The maximum products are obtained as $E \times P = 7.33\%$ and $E \times P = 8.04\%$ 
in the Ranking method with the PDE method and the BDT method respectively for $p=q=2$.

\begin{figure*}[!htbp]
\centering
\begin{center}
	\resizebox{0.6\textwidth}{!}{\includegraphics{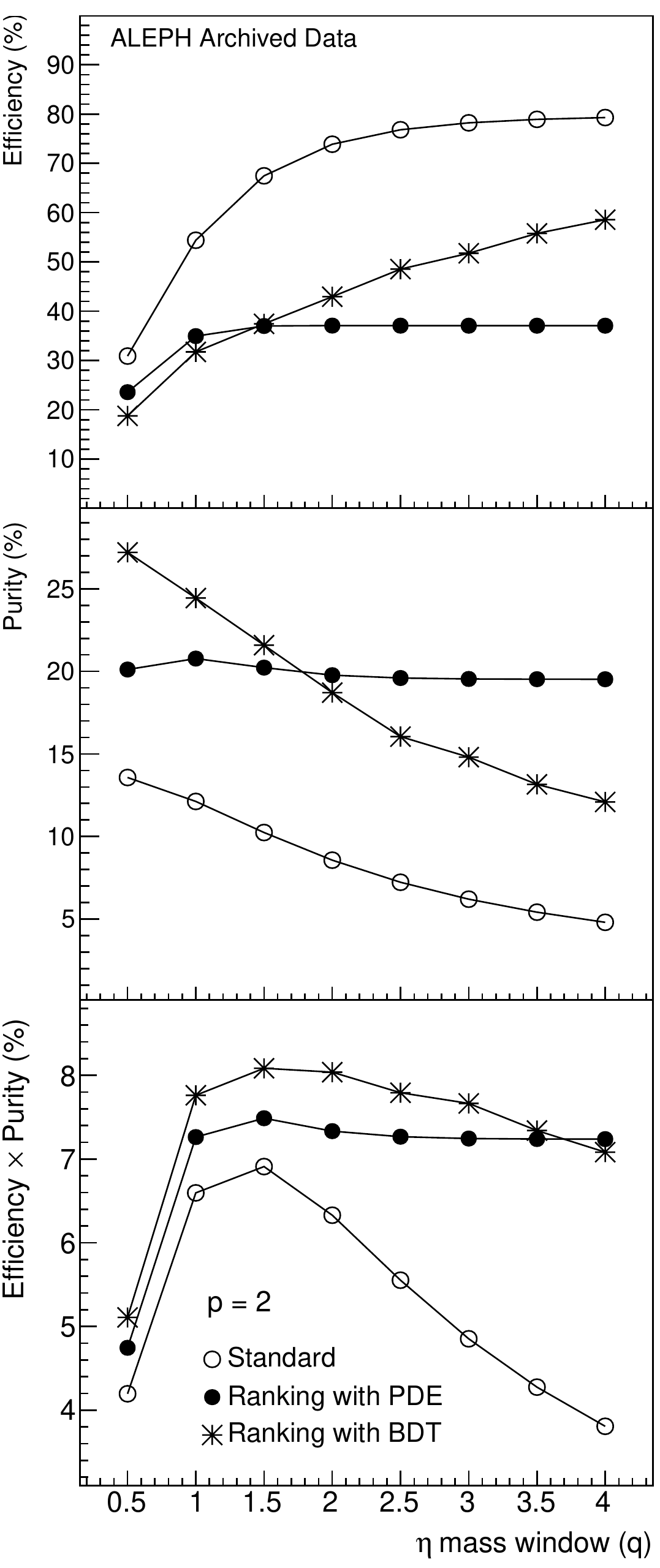}}
	\vspace{-0mm}
	\caption{\label{FIGURE_EFFPUR}
		Effect of varying $q$ on $\eta$ selection efficiency, purity and 
		their product values  
		when applying the standard method (open circles) and 
		the Ranking method with PDE (full circles) and with BDT (stars) 
		for the fixed value of $p=2$.
		The selection efficiency of $\eta$ signal candidates is calculated with respect to 
		mass window cuts corresponding to $p=0$ and $q=6$.
		The error bars at each point are much smaller than the marker sizes.
	}
\end{center}
\end{figure*}

%In fact, for all methods, optimum selections correspond to $p = 2$ and $q = 1.5$ which 
%are not shown in the table. 
Figure~\ref{FIGURE_EFFPUR} shows 
the efficiency, the purity and the product values as a function of $q$  
for the fixed value of $p=2$.
The Ranking method with PDE (full circles) and with BDT (stars) result in higher 
purity and product values with loss of efficiency relative to the 
standard method (open circles) at each point.
It is clear that Ranking with PDE exhibits a close performance to that of 
Ranking with BDT which appears to exhibit the best performance.

%%%%%%%%%%%%%%%%%%%%%%%%%%%%%%%%%%%%%%%%%%%%%%%%%%%%%%%%%%%%%%%%%%%%%%%%%%%%%%%%%%%%%%%
\section{Case study}
\label{SECTION_CASESTUDY}
$\eta$ selection methods described in the previous
sections are applied to the decay channel
$\eta' \rightarrow \pi^+ \pi^- \eta$, 
($BR = 42.9 \pm 0.7\%$)~\cite{CITE_PDG}.
In ALEPH, charged pions are measured in the tracking chamber
with a good momentum resolution.
However, the reconstructed momentum resolution of
$\eta \rightarrow \gamma \gamma$ decays 
is very poor since the energy resolution of the electromagnetic calorimeter
for unconverted photons is worse.
In order to improve the momentum resolution of $\eta$ candidates,
the reconstructed mass of photons is constrained to the nominal $\eta$ mass
using a fast method described in~\cite{CITE_MASCON}.

\begin{figure*}[!htbp]
\centering
\begin{center}
	\resizebox{0.8\textwidth}{!}{\includegraphics{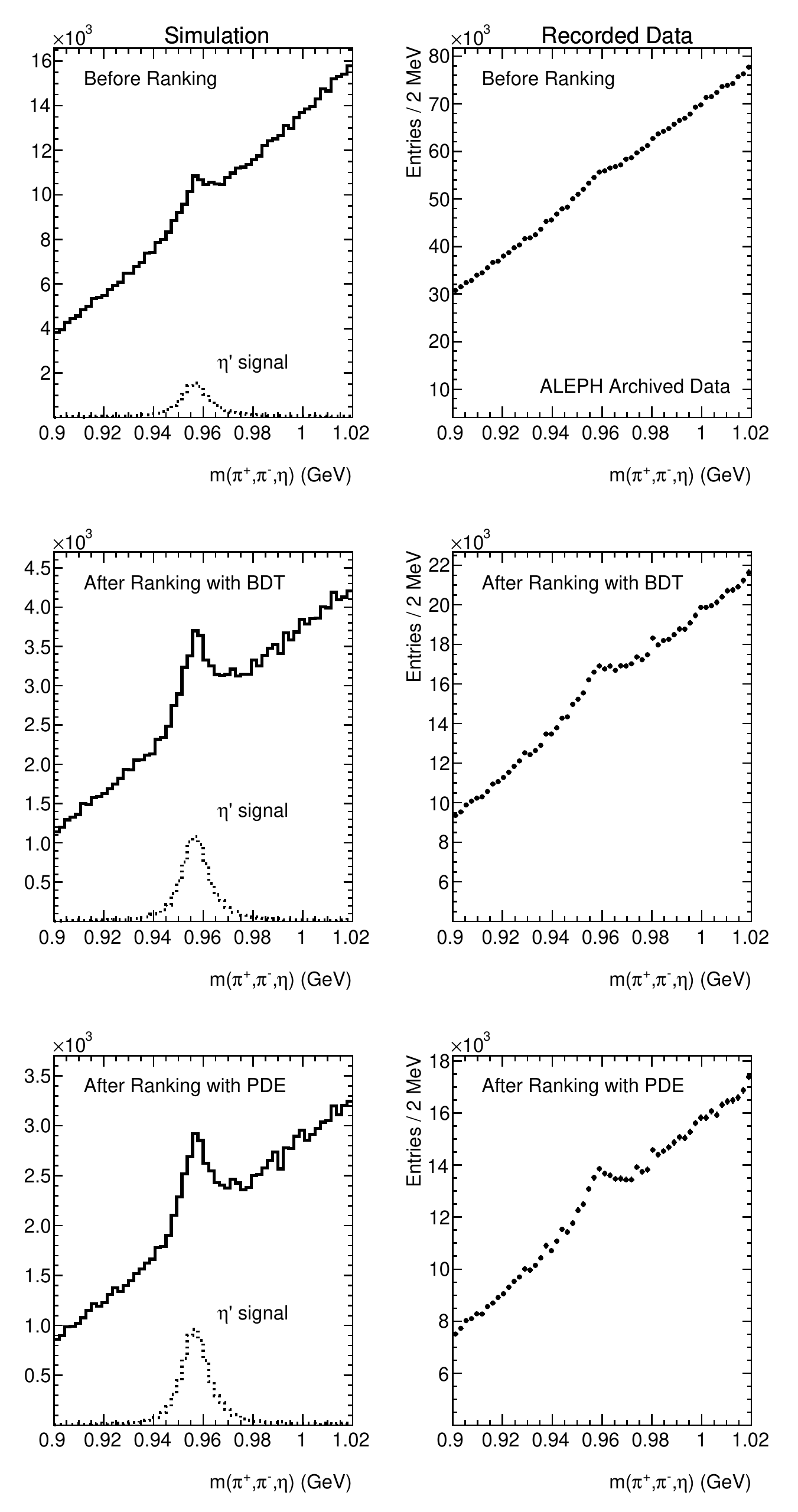}}
	\vspace{-0mm}
	\caption{\label{FIGURE_ETAPRIME}
		Invariant mass spectra of $\pi^+, \pi^-$ and $\eta$ candidates
		selected from hadronic $Z$ decays.
		An $\eta'$ signal is seen around 0.958 GeV.
		Before Ranking, the standard method is applied for $p = q = 2$.
		After Ranking (with PDE or with BDT), the signal appears much clearer 
		for both simulated (left column) and recorded ALEPH data (right column).
	}
\end{center}
\end{figure*}

Figure~\ref{FIGURE_ETAPRIME} shows 
example mass spectra for simulated and recorded ALEPH data.
Before Ranking, the standard method is applied
such that $\eta$ candidates are selected
for the case corresponding to\footnote{Although optimum 
	values are $p=2$ and $q=1.5$, we have used $p = q = 2$ to reduce 
	possible systematic uncertainties originating from mass window cuts.}
$p = q = 2$. Then, applying the Ranking method (with PDE and BDT estimators)
to remove some combinatorial background (for the same mass window cuts),
results in a clearer $\eta'$ signal around 0.958 GeV. 
Evidently, the Ranking methods improve the statistical significance of the $\eta'$ signal,
and can be expected to improve the fit stability when fitting mass spectra whose background
is very large.

As an alternative, we have also tried to use the PDE and 
BDT methods without Ranking. For this case, estimator cuts are optimized 
separately for the analysis with and without Ranking.
The performance, however, is found to be significantly worse. 

%%%%%%%%%%%%%%%%%%%%%%%%%%%%%%%%%%%%%%%%%%%%%%%%%%%%%%%%%%%%%%%%%%%%%%%%%%%%%%%%%%%%%%%
\section{Summary and conclusion}
\label{SECTION_CONCLUSION}
Reconstruction of $\eta$ mesons plays a dominant role
in the resultant purity and momentum resolution of reconstructed mother particles.
The purity of $\eta$ signals, for the decay channel $\eta \rightarrow \gamma \gamma$, 
is very poor due to the large combinatorial background in high multiplicity hadronic events.
In this study, two methods for the improvement of selection purity of $\eta$ mesons
are presented; the cut-based standard method and the Ranking method.
The Ranking method results in significant improvements
in $\eta$ purity in high multiplicity events.
However, it does not perform well if an event contains a small 
number of $\eta$ candidates (such as one or two).
In addition, the Ranking method using a PDE exhibits a close performance to that of 
the Ranking using a BDT which appears to exhibit the best performance.

The methods are applied in the reconstruction of the decay $\eta' \rightarrow \pi^+ \pi^- \eta$.
It is found that the selection of $\eta$ candidates 
with a higher purity using the Ranking method improves
the significance of the $\eta'$ signal relative to the standard selection method.

Finally, the authors suggest that the methods discussed here
can also be employed to extract short lived particles,
whose systematic errors are dominated by uncertainties
arising from the fitting procedure, such as
$J/\psi \rightarrow \pi^+ \pi^- \eta$ and $D^0 \rightarrow \eta \eta$
in proton-proton collision events at LHC 
where particle multiplicities are very high.

%%%%%%%%%%%%%%%%%%%%%%%%%%%%%%%%%%%%%%%%%%%%%%%%%%%%%%%%%%%%%%%%%%%%%%%%%%%%%%%%%%%%%%%
\section{Acknowledgement}
The authors wish to thank the ALEPH collaboration for access to the
archived data since the closure of the collaboration~\cite{CITE_ALEPH_RULE}.

%%%%%%%%%%%%%%%%%%%%%%%%%%%%%%%%%%%%%%%%%%%%%%%%%%%%%%%%%%%%%%%%%%%%%%%%%%%%%%%%%%%%%%%

%%%%%%%%%%%%%%%%%%%%%%%%%%%%%%%%%%%%%%%%%%%%%%%%%%%%%%%%%%%%%%%%%%%%%%%%%%%%%%%%%%%%%%%%%%%

%uncomment the following lines to place a figure
%\begin{figure}[htb]
%\centerline{%
%\includegraphics[width=12.5cm]{Fig1}}
%\caption{Plot of ...}
%\label{Fig:F2H}
%\end{figure}

\newpage

%%%%%%%%%%%%%%%%%%%%%%%%%%%%%%%%%%%%%%%%%%%%%%%%%%%%%%%%%%%%%%%%%%%%%%%%%%%%%%%%%%%%%%%
\begin{table}[H]
\caption{Effect of varying the mass window scales $p$ and $q$ on the selection 
	efficiency and purity of $\eta$ candidates using the standard method (STD) and 
	the Ranking method with PDE and with BDT. 
	The number of $\eta$ signal ($S$) and background ($B$) candidates are given as well.}
\smallskip
\label{TABLE_EFFPUR}
\centering
\setlength{\tabcolsep}{1mm}
\small
\renewcommand*{\arraystretch}{1.2}
%	\rowcolors{1}{}{gray!22}
\newcommand{\renk}{\rowcolor{gray!22}}
\begin{tabular}{lrrrrrrr}
	\hline
	Method & \multicolumn{1}{c}{\it p} & \multicolumn{1}{c}{\it q} & \multicolumn{1}{c}{\it S} & \multicolumn{1}{c}{\it B} & \multicolumn{1}{c}{\it E(\%)} & \multicolumn{1}{c}{\it P(\%)} & \multicolumn{1}{c}{$E\times P$} \\\hline
	STD & 0 & 1 & 222255 & 3321452 & 68.5 & 6.3 & 4.30\\\hline
	STD & 0 & 2 & 301772 & 6690397 & 93.0 & 4.3 & 4.01\\\hline
	STD & 0 & 3 & 318722 & 10147620 & 98.2 & 3.0 & 2.99\\\hline
	STD & 0 & 4 & 322604 & 13707263 & 99.4 & 2.3 & 2.29\\\hline
	\renk STD & 1 & 1 & 196960 & 1801165 & 60.7 & 9.9 & 5.98\\\hline
	\renk PDE & 1 & 1 & 125242 & 604917 & 38.6 & 17.2 & 6.62\\\hline
	\renk BDT & 1 & 1 & 110752 & 423415 & 34.1 & 20.7 & 7.08\\\hline
	STD & 1 & 2 & 267259 & 3611326 & 82.4 & 6.9 & 5.68\\\hline
	PDE & 1 & 2 & 132501 & 677472 & 40.8 & 16.4 & 6.68\\\hline
	BDT & 1 & 2 & 150675 & 815142 & 46.4 & 15.6 & 7.24\\\hline
	\renk STD & 1 & 3 & 282665 & 5434268 & 87.1 & 4.9 & 4.31\\\hline
	\renk PDE & 1 & 3 & 132430 & 685707 & 40.8 & 16.2 & 6.61\\\hline
	\renk BDT & 1 & 3 & 181834 & 1301457 & 56.0 & 12.3 & 6.87\\\hline
	STD & 1 & 4 & 286337 & 7256465 & 88.2 & 3.8 & 3.35\\\hline
	PDE & 1 & 4 & 132426 & 686326 & 40.8 & 16.2 & 6.60\\\hline
	BDT & 1 & 4 & 197262 & 1642840 & 60.8 & 10.7 & 6.52\\\hline
	\renk STD & 2 & 1 & 176513 & 1279482 & 54.4 & 12.1 & 6.60\\\hline
	\renk PDE & 2 & 1 & 113410 & 432452 & 35.0 & 20.8 & 7.26\\\hline
	\renk BDT & 2 & 1 & 103029 & 318512 & 31.8 & 24.4 & 7.76\\\hline
	STD & 2 & 2 & 239710 & 2558780 & 73.9 & 8.6 & 6.33\\\hline
	PDE & 2 & 2 & 120335 & 488306 & 37.1 & 19.8 & 7.33\\\hline
	BDT & 2 & 2 & 139364 & 605411 & 43.0 & 18.7 & 8.04\\\hline
	\renk STD & 2 & 3 & 253798 & 3837930 & 78.2 & 6.2 & 4.85\\\hline
	\renk PDE & 2 & 3 & 120277 & 495194 & 37.1 & 19.5 & 7.24\\\hline
	\renk BDT & 2 & 3 & 167943 & 966310 & 51.8 & 14.8 & 7.66\\\hline
	STD & 2 & 4 & 257260 & 5101272 & 79.3 & 4.8 & 3.81\\\hline
	PDE & 2 & 4 & 120274 & 495755 & 37.1 & 19.5 & 7.24\\\hline
	BDT & 2 & 4 & 190039 & 1381700 & 58.6 & 12.1 & 7.08\\\hline
	\renk STD & 3 & 1 & 160462 & 1074061 & 49.5 & 13.0 & 6.43\\\hline
	\renk PDE & 3 & 1 & 103425 & 368805 & 31.9 & 21.9 & 6.98\\\hline
	\renk BDT & 3 & 1 & 95637 & 271265 & 29.5 & 26.1 & 7.68\\\hline
	STD & 3 & 2 & 217820 & 2146755 & 67.1 & 9.2 & 6.18\\\hline
	PDE & 3 & 2 & 109868 & 418822 & 33.9 & 20.8 & 7.04\\\hline
	Bdt & 3 & 2 & 131311 & 534437 & 40.5 & 19.7 & 7.98\\\hline
	\renk STD & 3 & 3 & 230832 & 3216074 & 71.1 & 6.7 & 4.76\\\hline
	\renk PDE & 3 & 3 & 109814 & 425257 & 33.8 & 20.5 & 6.95\\\hline
	\renk BDT & 3 & 3 & 161957 & 898342 & 49.9 & 15.3 & 7.62\\\hline
	STD & 3 & 4 & 234102 & 4269366 & 72.1 & 5.2 & 3.75\\\hline
	PDE & 3 & 4 & 109811 & 425806 & 33.8 & 20.5 & 6.94\\\hline
	BDT & 3 & 4 & 176812 & 1144304 & 54.5 & 13.4 & 7.29\\\hline
\end{tabular}
\end{table}



\begin{thebibliography}{00}
\bibitem{CITE_PDG}
C. Patrignani et al, (Particle Data Group), {\em Chin. Phys. C, 40, 100001 (2016)} 
%---------------------------------------------
\bibitem{CITE_ALEPH_RULE}
Statement on use of ALEPH data for long-term analyses \\
(https://hep-project-dphep-portal.web.cern.ch/content/aleph-preservation-policy)
%---------------------------------------------
\bibitem{CITE_ALEPH}
D. Buskulic et al., (ALEPH Collab.), {\em Nucl. Instr. Meth. A360 (1995) 481}
%---------------------------------------------
\bibitem{CITE_ALEPH_EVENT}
D. Buskulic, et al. (ALEPH Collab.), {\em Z. Phys. C 69 (1996) 379}
%---------------------------------------------
\bibitem{CITE_ALEPH_ETA}
ALEPH Collaboration, {\em Eur. Phys. J C16 (2000) 597}
%---------------------------------------------
\bibitem{CITE_ROOT}
A. Hoecker, et al., {\em Proc. Sci., A CAT2007 (2007) 040}
[arXiv:physics/0703039]  
%---------------------------------------------
\bibitem{CITE_RANKING}
A. Beddall, et al., {\em Nucl. Instrum. Methods A 482 (2002) 520}
%---------------------------------------------
\bibitem{CITE_MASCON}
A. Bing\"ul, {\em Nucl. Instrum. Methods A 693 (2012) 11}
%---------------------------------------------
\end{thebibliography}
\end{document}